\documentclass{icrc2009}

\usepackage{graphicx}   
\usepackage[caption=false]{caption}    
\usepackage[font=footnotesize]{subfig} 
\usepackage{fixltx2e}
\usepackage{url}

\newcommand{\shorttitle}[1]%
{\markboth{Proceedings of the 31\MakeLowercase{$^{st}$} ICRC, {\L}\'{o}d\'{z} 2009}{#1} }
\newcommand{\etal}{\MakeLowercase{\textit{et al. }}} 

\begin{document}
\title{VERITAS Observations of Globular Clusters.}

\author{\IEEEauthorblockN{Michael McCutcheon\IEEEauthorrefmark{1} for the VERITAS Collaboration\IEEEauthorrefmark{2}}
	\\
	\IEEEauthorblockA{\IEEEauthorrefmark{1} Department of Physics, McGill University, Montreal, QC, H3A 2T8, Canada (mccutchm@physics.mcgill.ca)}
	\IEEEauthorblockA{\IEEEauthorrefmark{2} see R.A. Ong \etal (these proceedings) or \textsf{http://veritas.sao.arizona.edu/conferences/authors?icrc2009}.} } 

\shorttitle{M. McCutcheon \etal -  Globular Clusters with VERITAS.}
\maketitle

\begin{abstract}
It has been postulated that globular clusters could be sources of Very-High Energy (VHE) gamma rays, powered by milli-second pulsars. 
This could be due to cumulative direct emission or to plerion-type emission driven by colliding winds. 
In particular the southern hemisphere globular cluster 47 Tuc has been singled out as a potential source in both models. 
In light of the recent detection by the Fermi Gamma-ray Space Telescope (FGST) of 47 Tuc, the first detection of any globular cluster as a gamma-ray source, we present the results of observations of northern hemisphere globular clusters by VERITAS. 
Three globular clusters have been observed: M15, M13 and M5. Of these, M15 and M13 have been explicitly proposed as VHE gamma-ray sources and M5 possess similarities with them.

\end{abstract}

 

  \begin{table*}[t]
  \caption{Globular clusters containing millisecond pulsars and observed by VERITAS.}
  \label{target_table}
  \centering
  \begin{tabular}{|c|c|c|c|c|c|c|}
  \hline
  Object & \# known pulsars	& Distance (kpc) & $m_\textrm{V}$  & $L$ ($10^{3} L_\odot$) &  $U_\textrm{rad}$  ($\textrm{eV} / \textrm{cm}^{3}$) & $R_C$  (pc) \\
  \hline
  M15	& 8	& 10.2	& 6.20 & 365	& 5035	& 0.21	\\
  M13	& 5	& 7	& 5.78 & 208	& 131	& 1.79	\\
  M5	& 5	& 7.3	& 5.65 & 260	& 463	& 0.85	\\
  \hline
  \end{tabular}
  \end{table*}

  \begin{table*}[t]
  \caption{Count rates and flux upper limits from VERITAS data.}
  \label{summary_table}
  \centering
  \begin{tabular}{|c|c|c|c|c|c|c|c|c|}
  \hline
  Object & \# Telescopes & Exposure & $\textrm{N}_\textrm{ON}$ & $\textrm{N}_\textrm{OFF}$ & $\alpha$ & Significance & \multicolumn{2}{|c|}{Flux Upper Limit ($E>600$ GeV)} \\
  	& 		& (min)	& 	& 		& 	& ($\sigma$) & ($10^{-12}$ erg/s/cm$^2$) & (\% Crab) \\
  \hline
  M15	& 2	& 393	& 10	& 61	& 0.25	& -1.3	& 1.1 & 1.6	\\
  M13	& 3-4	& 397	& 12	& 87	& 0.10	& 1.0	& 1.5 & 2.2	\\
  M5	& 4	& 900 	& 25	& 251	& 0.11	& -0.3	& 0.4 & 0.6	\\
  \hline
  \end{tabular}
  \end{table*}

\section{Introduction}

Globular clusters are ancient, dense clusters of stars 
for which the central stellar density can exceed that of the solar neighborhood by a factor of 500. 
It has been proposed 
\cite{SpinUp}
that such dense stellar environments may directly lead to 
the formation of large numbers of millisecond radio pulsars (MSPs); the latter a result of spinning-up by the 
accretion of matter from low-mass stellar companions, 
which can be acquired by binary-exchange interactions.
Quantitative analysis of this process (i.e. \cite{mspContent}, \cite{mspPop}) 
predicts the existence of $10-100$ MSPs in individual, large globular clusters.
Surveys of globular clusters have discovered 140 radio pulsars with spin periods in the range of a few 
to a few tens of milliseconds in 26 globular clusters: 
e.g. \cite{mspReview}. 
However these searches are at present completely sensitivity limited, 
such that it is expected that only the brightest pulsars have yet been observed 
and estimates of the total pulsar population vary up to $10^4$ \cite{mspContent}.

Given a large, dense population of MSPs within a globular cluster, two feasible mechanisms for 
detectable emission of Very High Energy (VHE) gamma rays are :
\begin{enumerate}
\item The combined direct emission from particles accelerated in the magnetospheres of individual pulsars \cite{mspGamma}.
\item Inverse Compton scattering of ambient stellar photons by particles accelerated in shocks formed by colliding pulsar winds \cite{gc_mspWinds}.
\end{enumerate}

The gamma-ray spectra determined in \cite{mspGamma} for MSPs, using the polar cap emission model, suggest a curvature radiation peak at $\sim25$ GeV next to an exponential cutoff, effectively eliminating emission beyond 100 GeV.  
An Inverse Compton (IC) component exists from the scattering of thermal X-ray photons near the heated polar caps, but this is strongly absorbed by the pulsar's magnetic field and, for an individual pulsar, remains several orders of magnitude below the current sensitivity of Imaging Air Cherenkov Telescopes (IACTs).   
Considering that the combined emission of a globular cluster's MSP population would reflect this overall pattern, this emission mechanism disfavours detection by IACTs in contrast to FGST.

The emission mechanism involving colliding pulsar winds, as detailed in \cite{gc_mspWinds}, relies on the IC scattering of the intense optical photon field in the core of the globular cluster by shock-accelerated electrons. 
It is stated in \cite{gc_mspWinds} that because VHE gamma-ray emission in this scenario is by the IC channel, the spectrum favours detection by IACTs over detectors operating at a lower threshold, due to the peaks of a number of the possible spectra being placed in the 0.1 - 1 TeV range.

VERITAS has observed three globular clusters which are reasonable candidates for these emission mechanisms :  M15, M13 and M5.  
Table~\ref{target_table} lists selected characteristics of these objects: 
$m_\textrm{V}$ is the visible magnitude; 
$L$ is the total luminosity; 
$U_\textrm{rad}$ is the central radiation density; 
$R_C$ is the core radius
(see \cite{GCtable} for more details).
Each harbours a significant number of known pulsars, has an intense central radiation field and all are expected to have large stellar interaction rates. 

Analysis results for these observations are presented and upper limits for gamma-ray flux derived.  Given the variety of spectra which could be produced, according to (and within) the specific models of emission considered here, for simplicity a Crab Nebula-like spectrum (power law with index -2.5) is assumed.

\section{Analysis \& Results}

VERITAS is a gamma-ray observatory using the imaging air Cherenkov technique \cite{Vtech}.  
It is situated in Arizona, USA, and consists of four, 12-m Davies-Cotton telescopes.  
It is currently the most sensitive IACT in the Northern hemisphere; capable of detecting a source with a flux of 10 milli-Crab in less than 50 hours of observations.

Data used in this analysis were taken both during VERITAS's commissioning phase and subsequent full operation.

To assure data quality, observation periods were rejected from the analysis if they exhibited a sub-nominal trigger rate (for the appropriate elevation) and/or evidence of clouds (in readings from a co-pointing FIR pyrometer).

Images of extensive air showers are extracted from the data by requiring that the charge in individual pixels exceeds the pedestal variance due to night-sky background photons by a certain ratio.
In addition, the images must pass quality cuts before being subject to second-moment parametrization and subsequent stereo reconstruction.
Specifically, for this analysis, the image quality cuts include the requirements that:
\begin{itemize}
\item[-] an image must contain at least 4 contiguous pixels.
\item[-] less than 20\% of an image's total charge can come from pixels at the edge of the camera.
\item[-] the total charge contained in an image must exceed that equivalent to 100 photo-electrons.  
\end{itemize}
After stereo reconstruction, further quality and gamma/hadron separation cuts are applied. 

The rate of gamma rays from the direction of the target may then be determined.
The ON region is defined by $\theta^2 < 0.015~\textrm{deg}^2$ (where $\theta$ is the angular separation from the target) and the background rate is determined using the reflected region model.  
The significance has been calculated according to equation 17 of \cite{LiMa} and the 99\% confidence-level upper limits presented here were calculated using the method of Helene \cite{Helene}.

For the M13 and M5 data sets, taken since VERITAS's official first-light, effective areas were determined from simulations appropriate to the complete detector layout.  
The M15 data set was collected during the commissioning phase, for which appropriate simulations are not readily available :  in this case the flux has been calculated by scaling to contemporaneous data taken on the Crab Nebula in the same elevation band. 
The analysis results are shown in Table~\ref{summary_table}.


\section{Discussion \& Conclusions}

No VHE gamma-ray emission was detected above 600 GeV from these three globular clusters.

Given that the mechanism of cumulative direct emission does not call for a significant flux above 600 GeV 
this result leaves it entirely posible that, for instance, FGST will be able to detect emission from these objects via this mechanism.

On the other hand, these limits do allow us to speak to the assumptions of the colliding-winds model.  
Assuming that the details of the colliding-winds shock fronts yield a spectrum reasonably similar to the power law assumed for these limits, we can estimate the number of pulsars allowed by these limits in conjunction with the model.  
In order to derive spectra for specific globular clusters (M15 and M13), it is assumed in \cite{gc_mspWinds} that each cluster contains a population of $N_p = 100$ MSPs, each with an electron acceleration efficiency of $\eta = $1\% of the typical spin-down luminosity and that the flux will scale as $N_p \eta$. 
Comparing the limits here to an estimate of the most optimistic fluxes near 1 TeV from \cite{gc_mspWinds} we can simply scale the pulsar population appropriately for M15 and M13.
For M5 we first have to estimate how it compares to the predictions for M15 and M13.  A conservative assumption is that the flux for this mechanism scales according to:
\begin{displaymath}
	\frac{ U_{\textrm{rad}} R_{c}^{3} }{ d^2 }
\end{displaymath}
where $d$ is the distance of the globular cluster.
Furthermore, the characterisitics of M5 are more similar to M13 than to M15; in particular M15 is a core-collapse globular cluster, in a later stage of gravitational evolution.  
Thus, the flux estimate for M5 is scaled to that for M13.
The limit on the pulsar population for each globular cluster is indicated in Table~\ref{limits_table}.

  \begin{table}[!t]
  \caption{Pulsar population limits within the colliding-winds model considering the calculated upper limits.}
  \label{limits_table}
  \centering
  \begin{tabular}{|c|c|c|}
  \hline
  Object & Estimated Flux for $N_p \eta = 1 $  	& Pulsar Population Limit	\\
  	& (\% Crab @ 1 TeV)			& for $\eta = 0.01$		\\
  \hline
  M15	& 3.1	& 53	\\
  M13	& 6.2	& 36	\\
  M5	& 2.1	& 30 	\\
  \hline
  \end{tabular}
  \end{table}

We can conclude that, within the context of the most optimistic version of the colliding-winds mechanism as depicted in \cite{gc_mspWinds}, the pulsar population of these globular clusters can be estimated as being within an order of magnitude of the presently known pulsar content.

\section{Acknowledgements}
This research is supported by grants from the US Department of Energy, the US National Science Foundation, and the Smithsonian Institution, by NSERC in Canada, by Science Foundation Ireland, and by STFC in the UK. We acknowledge the excellent work of the technical support staff at the FLWO and the collaborating institutions in the construction and operation of the instrument.\\[128ex]


\end{document}